\documentclass[10pt,journal,compsoc]{IEEEtran}
\ifCLASSOPTIONcompsoc
  \usepackage[nocompress]{cite}
\else
  \usepackage{cite}
\fi
\usepackage{amsmath,amsfonts}
\usepackage{algorithmic}
\usepackage{algorithm}
\usepackage{array}
\usepackage{textcomp}
\usepackage{stfloats}
\usepackage{url}
\usepackage{verbatim}
\usepackage{graphicx}
\usepackage{cite}
\usepackage{pbox}
\usepackage{xcolor}
\usepackage{hyperref}
\usepackage{listings}
\usepackage{caption}
\usepackage{subcaption}
\usepackage{comment}
\newtheorem{definition}{Definition}
\usepackage{lscape}
\def\BibTeX{{\rm B\kern-.05em{\sc i\kern-.025em b}\kern-.08em
    T\kern-.1667em\lower.7ex\hbox{E}\kern-.125emX}}

\begin{document}

\title{A Survey of MulVAL Extensions and Their Attack Scenarios Coverage}

\author{\IEEEauthorblockN{David~Tayouri, Nick~Baum, Asaf~Shabtai, Rami~Puzis}\\
\IEEEauthorblockA{\textit{Dept. of Software and Information Systems Engineering} \\
\textit{Ben-Gurion University of the Negev}\\
Beer-Sheva, Israel \\
\{davidtay,nicita\}@post.bgu.ac.il, \{shabtaia,puzis\}@bgu.ac.il}
}

\IEEEtitleabstractindextext{%
\begin{abstract}
Organizations employ various adversary models in order to assess the risk and potential impact of attacks on their networks.
Attack graphs represent vulnerabilities and actions an attacker can take to identify and compromise an organization's assets.
Attack graphs facilitate both visual presentation and algorithmic analysis of attack scenarios in the form of attack paths. 
MulVAL is a generic open-source framework for constructing logical attack graphs, which has been widely used by researchers and practitioners and extended by them with additional attack scenarios.
This paper surveys all of the existing MulVAL extensions, and maps all MulVAL interaction rules to MITRE ATT\&CK Techniques to estimate their attack scenarios coverage.
This survey aligns current MulVAL extensions along unified ontological concepts and highlights the existing gaps.
It paves the way for methodical improvement of MulVAL and the comprehensive modeling of the entire landscape of adversarial behaviors captured in MITRE ATT\&CK.   
\end{abstract}

}

\maketitle
\IEEEpeerreviewmaketitle

\IEEEraisesectionheading{\section{Introduction}}
With the growth in the number of cyber attacks and their increasing complexity, cyber security risk assessment has become more essential~\cite{johns2020cyber, furnell2020understanding}.
To improve their cyber security, organizations must identify their business critical elements and protect them.
For every possible threat, there may be a number of countermeasures; since it is infeasible to implement all countermeasures, organizations should assess the risks to their systems, prioritize these risks, and identify the security measures that will best reduce the threats to an acceptable level~\cite{landoll2005security}.

Different attack modeling techniques can be used to perform risk assessment and present the risks visually, including misuse sequence diagrams (a use case method)~\cite{katta2010comparing}, cyber kill-chain (a temporal method)~\cite{hutchins2011intelligence}, and fault trees~(a graph-based method)~\cite{haasl1981fault}.
A popular method of visually representing cyber risks is the attack graph.
An attack graph is a risk assessment method aimed at representing attack states, transitions between them, and the related enterprise network vulnerabilities~\cite{phillips1998graph}.
Attack graphs organize identified vulnerabilities into attack paths, composed of sequences of actions an attacker can take in order to reach and compromise system assets. 
Attack graphs can also help identifying the attack paths most likely to succeed.
As a consequence, attack graphs enable security administrators to prioritize the risks to an organization's network and decide which vulnerabilities to patch first.

Most attack graphs suffer from scalability challenges when modeling large networks~\cite{sabur2022toward}.
Some frameworks address these challenges by adding assumption such as the delete-free relaxation in logical attack graphs~\cite{ou2006scalable}.  
Nevertheless, attack graphs have two main advantages over other risk assessment methods.
First, an attack graph models the interactions between vulnerabilities (multi-stage attacks) and the attacker's lateral movements (multi-host attacks), instead of focusing on individual vulnerabilities. 
Second, for the preconditions, consequences, and severity, attack graph risk assessment considers the effect of the exploitation of vulnerabilities on the specific target environment.

Different types of attack graphs have been proposed, including attack trees~\cite{schneier1999attack}, state graphs~\cite{sheyner2002automated}, exploit dependency graphs~\cite{noel2003efficient}, logical attack graphs~\cite{ou2006scalable}, and multiple prerequisite attack graphs~\cite{ingols2006practical} (a brief overview of attack graphs is presented in Section~\ref{AttackGraphs}).
In this research, we focus on the logical attack graph - a directed graph in which leaves represent facts about the system, the internal nodes represent actions (attack steps) and their consequences (privileges), and the root represents an attacker’s final goal.
MulVAL is a well-known open-source framework for constructing logical attack graphs~\cite{mulval}.
In addition to its scalability and extensibility, MulVAL is commonly used by researchers; as of February 2022, we were able to identify 938 academic publications that mention MulVAL.
A description of the MulVAL framework is presented in Section~\ref{MulvalFramework}.

To generate an attack graph, MulVAL requires four main inputs: security domain knowledge, such as CVE (Common Vulnerabilities \& Exposures); information regarding the environment state, such as the principals, and network and host configuration; the security policy; and reasoning rules.
MulVAL's reasoning engine relies on interaction rules, which describe how facts and privileges are used by actions to achieve attack goals.
The original MulVAL framework provided a set of interaction rules that represented a limited set of attacks.
Since MulVAL was introduced in 2005, interaction rules have been added to represent additional attack scenarios.
Researchers interested in using all MulVAL interaction rules would need to perform a comprehensive review of the literature and search through the hundreds of papers that mention MulVAL.
Our first goal was to review all of these papers in order to collect all of the additional MulVAL interaction rules.

To identify all of the academic publications presenting MulVAL extensions, we performed a systematic literature review. Of the 938 papers we identified (see Section~\ref{SearchMethod}), 38 extended MulVAL with additional interaction rules (see Section~\ref{ExtensionFindings}).
We provide a list of all of the MulVAL interaction rules we found in the literature (which are referred to as MulVAL rules in this paper)\footnote{The list of MulVAL rules is available at: \url{https://github.com/dtayouri/MulVAL-MITRE/blob/main/\%E2\%80\%8F\%E2\%80\%8FMulVAL\%20Interaction\%20Rules.xlsx}}.
Using the entire list of rules will enable the generation of attack graphs that cover more attack scenarios.

To provide a comprehensive assessment of the risks faced by an organization's network, attack graphs should be able to present as many attack scenarios as possible.
Thus, our second goal was to evaluate the extent to which the current MulVAL extensions cover known attack scenarios. 
The comprehensiveness and completeness of a set of interaction rules can be assessed using a knowledge base of known tactics, techniques, and procedures (TTPs).
MITRE ATT\&CK~\cite{mitreattackwebsite}, which is the defacto standard of cyber threat modeling taxonomies, is a globally-accessible evidence-based knowledge base of TTPs; MITRE ATT\&CK is described in Section~\ref{MitreAttack}. 
To evaluate the extent to which the MulVAL rules are able to represent different attacks, we systematically mapped all of the MulVAL rules to MITRE ATT\&CK Techniques.

Another important benefit of mapping all of the MulVAL rules to ATT\&CK Techniques is that mapping enables actionable insights: Techniques' Detection and Mitigations can be used to detect and mitigate the risks represented by the attack paths built with MulVAL rules.

\begin{figure}[htb]
    \centering
    \includegraphics[width=0.48\textwidth]{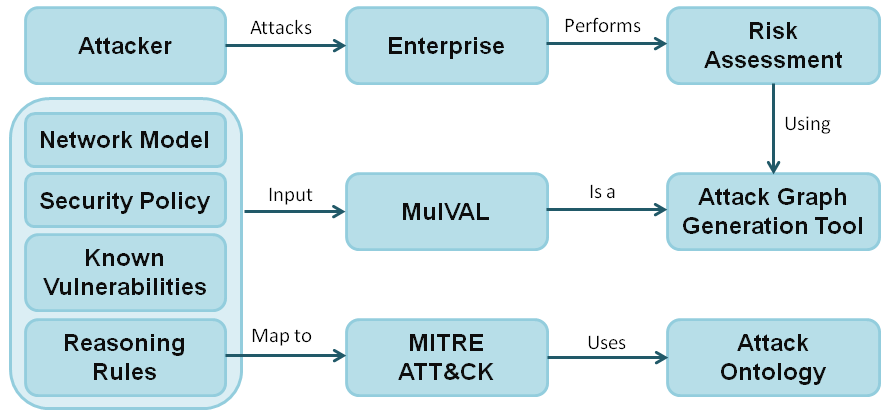}
    \caption{Relationships among the entities comprising the enterprise cyber ecosystem. }
    \label{fig:EntitiesRelationships}
\end{figure}

Figure~\ref{fig:EntitiesRelationships} presents the relationships between the different entities of the enterprise cyber ecosystem: attackers try to attack enterprise networks; enterprises perform risk assessment to prioritize the risks and allocate the resources to handle them; risk assessment can be performed by using an attack graph generation tool, such as MulVAL, for which there are several inputs; among the inputs are reasoning rules, which can be mapped to MITRE ATT\&CK to disclose the coverage of TTPs and enable the coverage of more TTPs.

The contributions of this work are:
\begin{itemize}
    \item We survey all of the MulVAL extensions found in the literature and provide the list of all published MulVAL rules.
    \item We map all available MulVAL rules to MITRE ATT\&CK Techniques and summarize the attack coverage capabilities of existing MulVAL extensions.
\end{itemize}

\section{Attack Graphs} \label{AttackGraphs}

\begin{figure*}[htb]
    \centering
    \includegraphics[width=0.95\textwidth]{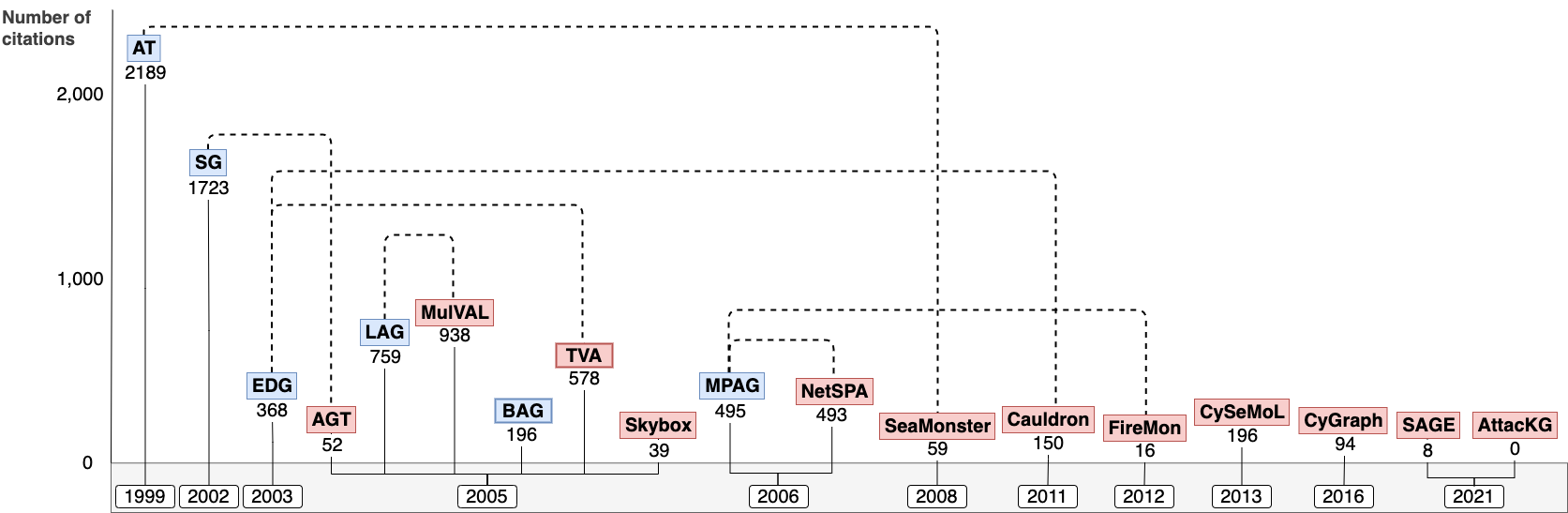}
    \caption{Evolution of attack graph generation methods: the publication year and number of citations of AG representations (blue) and the respective AG generation tools (red).}
    \label{fig:AGTimeline}
    %\Description{Attack graphs with years.}
\end{figure*}

% Section overview
An attack graph (AG) is a model that enables researchers and security administrators to provide a visual representation of events that may lead to a successful attack scenario.
Various AGs have been proposed in prior research.
Hong et al.~\cite{hong2017survey} conducted a survey reviewing all of the modeling techniques and AG generation tools presented in the literature.
In this section, we describe the most common AG representations, including the attack tree (AT), state graph (SG), exploit dependency graph (EDG), logical attack graph (LAG), and multiple prerequisite attack graph (MPAG) representations.
Figure~\ref{fig:AGTimeline} depicts these representations (in blue) and their supported AG generation tools (in red) on a timeline graph, along with the number of citations (y-axis).
We also review the common uses of AGs and the main challenges of modeling attacks with an AG.

% Attack graph common usages
The need for different AG representations stems from their use in diverse cyber domains and applications. 
For example, AG-based network security assessment methods can be utilized by modeling zero-day network resilience in an AG by defining a new zero-day safety metric that counts how many unknown vulnerabilities would be required to compromise network assets~\cite{wang2013k}.
Noel et al.~\cite{noel2008optimal} used AGs to solve the sensor-placement problem, optimally placing IDS (intrusion detection system) sensors by covering the entire AG using the fewest number of sensors.
Roschke et al.~\cite{roschke2013high} presented an AG-based IDS (an alert correlation algorithm capable of analyzing dependencies between vulnerabilities) aggregating similar alerts and deciding if an isolated alert is a part of an ongoing multi-step attack.
Liu et al.~\cite{liu2012using} used AGs for forensic analysis, by showing that security administrators can prove, for example, that a series of IDS alerts are not isolated, but rather correspond to a sequence of attacks in a coherent attack scenario.
Wang et al.~\cite{wang2006minimum} used AGs to solve the minimum network hardening problem by constructing the set of specific vulnerabilities that should be patched to eliminate the attack paths leading to a given critical asset, while minimizing the cost involved in removing those vulnerabilities.

% AT (Attack Tree) - [Tools: N/A]
\paragraph{Attack Tree (AT)} \label{AttackTree}
Tree-based graphical attack models are widely used to model network security~\cite{horne2017semantics, kumar2017quantitative, fila2020exploiting, nishihara2020validating}.
The most known tree-based representation, first published in 1999 by Schneier~\cite{schneier1999attack}, was the AT. 
The root node of an AT represents the attacker's goal, and leaf nodes represent the attacker's sub-goals.
Although, an AT does not enumerate all possible system states, it still depends on the number of events.
As a result, its main disadvantage is its poor scalability. 
In addition, modification of the AT nodes near the root node may result in modification of the entire tree.
SeaMonster is a commonly used open-source AT generation tool based on the Eclipse framework~\cite{meland2008seamonster}.
SeaMonster focuses on helping developers during the software development lifecycle by providing three different viewpoints: existing vulnerabilities in the software, what causes the vulnerabilities, and possible countermeasures.

% State Graph - [Tools: Attack Graph Toolkit]
\paragraph{State Graph (SG)} \label{StateGraph}
In 2002, Sheyner et al.~\cite{sheyner2002automated} presented the Attack Graph Toolkit, which is based on a SG.
The Attack Graph Toolkit utilizes the SG in which each node represents a global state of the network, and edges correspond to attack actions initiated by the intruder.
State enumeration-based approaches for AG representation suffer from degraded scalability. 
Enumerating all of the possible attack scenarios means dealing with a large state and action space, representing each possible system state as a node and each change of state caused by a single action taken by the attacker as an edge, resulting in a state space explosion~\cite{swiler1998graph}.
First introduced in 2002, Ammann et al.~\cite{ammann2002scalable} proposed a more scalable approach for AG representation called the monotonicity assumption.
The authors addressed the scalability issue by assuming that the preconditions of an attack are not invalidated by the successful execution of another attack.
Applying this assumption reduces the AG generation complexity from the exponential state space to the polynomial.

% EDG (Exploit Dependency Graph) - [Tools: TVA, Cauldron]
\paragraph{Exploit Dependency Graph (EDG)} \label{EDG}
In 2003, Noel et al.~\cite{noel2003efficient} presented the EDG, which enumerates all of the possible exploit sequences while considering the monotonicity assumption.
Each exploit or dependency represented appears only once, and all of the exploits contribute to the attack goal.
As a result, there are no edges between independent exploits, and the AG size is quadratic to the number of exploits.
However, enumerating all of the possible states of the attack using EDG is still an exponentially complex task.
To address this limitation, a heuristic method can be used.
In 2005, Jajodia et al.~\cite{jajodia2005topological} proposed the Topological Vulnerability Analysis (TVA) tool which is based on EDG. 
This tool uses two types of nodes: exploit and security condition nodes. 
Exploit nodes represent attack actions, and condition nodes represent either attack pre-conditions or post-conditions.
The graph is built backwards from the attacker goal to the initial exploit. 
As a result, they do not include exploits generated in the forward dependency graph, and all of the exploits are relevant to the predefined attack goal.
In addition, there is an enterprise version of TVA called Cauldron, which provides additional visualizations, data integration features, automatic generation of mitigation recommendations, etc.~\cite{cauldron}.

% LAG (Logical Attack Graph) - [Tools: MulVAL]
\paragraph{Logical Attack Graph (LAG)} \label{LAG}
In 2005, Ou et al.~\cite{ou2006scalable} introduced the LAG, a directed graph, which can also be represented as a tree.
Due to the monotonicity assumption, the size of the LAG is polynomial in the size of the network being analyzed.
A LAG can be generated using MulVAL's AG generation tool~\cite{ou2005logic}.
A description of this AG is provided in Section~\ref{MulvalFramework}.

% BAG (Bayesian Attack Graph)
\paragraph{Bayesian Attack Graph (BAG)} \label{BAG}
First proposed by Liu and Man in 2005~\cite{liu2005network}, the BAG is a directed acyclic graphical model where the nodes represent different security states that an attacker can acquire and the directed edges represent the dependencies between these security states.
The potential attack paths are modeled by assigning conditional probability tables to edges, enabling the use of Bayesian inference methods.
There is no generation tool available for BAG.

% MPAG (Multiple Prerequisite Attack Graph) - [Tools: NetSPA, FireMon]
\paragraph{Multiple Prerequisite Attack Graph (MPAG)} \label{MPAG}
In 2006, Ingols et al.~\cite{ingols2006practical} presented the MPAG.
%with additional way to reduce the number of nodes in an attack graph. 
The MPAG uses three types of nodes: state nodes, prerequisite nodes, and vulnerability nodes.
State nodes describe the attacker's level of access on a specific host, prerequisite nodes can represent the reachability group or a set of credentials, and vulnerability nodes express a particular vulnerability on a specific host.
MPAG node aggregation reduces the number of edges compared to a method in which state nodes point directly at vulnerability instance nodes, since many state nodes can imply the same set of attacks.
There are several AG generation tools that use MPAGs, such as NetSPA (network security planning architecture)~\cite{artz2002netspa} and FireMon~\cite{firemon}, which is a commercial attack generation tool based on NetSPA.
Both tools provide useful functionalities for security administrators, such as AG security assessment, prioritization of the vulnerabilities found, and suggestions on how to deal with the weaknesses discovered, however these tools also have some limitations.
For example, as an MPAG has many loops, this type of AG is difficult to understand.

In addition to the types of AGs mentioned above, there are some commonly used AG generation tools worth mentioning.

% (N/A) - [Tools: Skybox]
\paragraph{Skybox}
In 2005, Skybox View was presented by Skybox Security\footnote{\url{https://www.skyboxsecurity.com/}; the name of the tool is changed to Vulnerability Control} as a solution for vulnerability and threat management. 
As Skybox View is not an open-source product, its underlying AG representation is not publicly available.
However, like other commercial AG generation tools, it provides organizations with an end-to-end automated vulnerability management workflow and vulnerability discovery, assessment, prioritization, and remediation.

% Probabilistic relational model - [Tools: CySeMoL]
\paragraph{CySeMol}
In 2013, Holm et al.~\cite{holm2013cysemol, holm2013manual} presented the cyber security modeling language (CySeMoL), which is a modeling language and AG tool that can be used to estimate the cyber security of enterprise architectures.
CySeMoL includes theoretical information on how attacks and defenses relate quantitatively; thus, security expertise is not required of its users.
Users only need to model their system architecture and specify its characteristics in order to enable calculations.

% MPAG (Probabilistic relational model - CyGraph) - [Tools: CyGraph]
\paragraph{CyGraph}
In 2016, MITRE presented CyGraph, which is a graph-based AG generation tool~\cite{noel2016cygraph}.
This four-layer tool uses TVA/Cauldron as its network infrastructure and security posture layers. 
These layers import network topology information and search for vulnerabilities that might be exploited in cyber attacks. 
The other layers are cyber threats and mission dependencies, which are responsible for describing the potential cyber threats and capturing dependencies among various mission components.

\paragraph{CTI-Based Attack Graph}
In 2021, Nadeem et al.~\cite{nadeem2021alert} presented SAGE, a framework for constructing AGs from cyber threat intelligence (CTI), instead of system vulnerabilities. 
In the same year, Li et al.~\cite{li2021attackg} presented AttacKG, a method for extracting structured AGs from CTI reports and identifying the attack techniques.

% Summery of AG review
While we have described the most common AG representations and generation tools, there are also other kinds of AGs, such as DeepAG~\cite{li2022deepag}, which integrates AGs with deep learning techniques.

% AG challenges: visualization, scalability issues
The main challenges in modeling an AG are visualization and scalability.
Recently, Lallie et al.~\cite{lallie2020review} conducted a survey of 180 graphical attack representations proposed in the literature and concluded that more research is needed to standardize the representations. 
%As the amount of hosts in the organization network grows, it is crucial for the AG to be scalable.
The scalability of each AG type is reviewed as part of Table~\ref{table:AttackGraphTools}, which provides a comparison of the attack generation tools described above.
In this research, we focus on the MulVAL framework, which uses a logical attack graph and will be described in Section~\ref{MulvalFramework}.

\begin{table*}[h!tb]
    \centering
    \scriptsize
%    \tiny
    \caption{Comparison of common attack graph generation and visualization tools}
    \begin{tabular}{| m{5em} | m{8em} | m{5em} | m{5em} | m{5em} | m{4em} | m{2em} | m{3em} | m{10em} | m{7em} |} 

    \hline
    \textbf{Name} & \textbf{Developers} & \textbf{Accessible} & \textbf{AG Type} & \textbf{Scalability} & \textbf{Intuitive Level} & \textbf{Year} & \textbf{No. of References} & \textbf{Paper Search} & \textbf{Tool Search}\\ [0.5ex] 
    \hline
    Attack Graph Toolkit & \flushleft{Carnegie Mellon University} & Open source & \hyperlink{StateGraph}{SG} & Poor, Exponential & Fair & 2005 & 52 & \flushleft{"Scenario graphs applied to security": 16} & ["Attack Graph Toolkit"]: 52 \\ 
    \hline
    MulVAL & \flushleft{Kansas State University} & Open source & \hyperlink{LAG}{LAG} & O($N^2$) - O($N^3$) & Good & 2005 & 938 & "A scalable approach to attack graph generation": 757 & ["MulVAL"]: 938 \\
    \hline
    TVA & \flushleft{George Mason University} & \flushleft{Not open source, difficult to obtain} & \hyperlink{EDG}{EDG} & O($N^3$) & Good & 2005 & 578 & "Topological analysis of network attack vulnerability": 578 & ["Topological Vulnerability Analysis"]: 547 \\
    \hline
    Skybox View & \flushleft{Skybox Security, Inc.} & Commercial Software & Unknown & O($N^3$) & Good & 2005 & 39 & "Proactive Security for a Mega-Merger": 39 & ["skybox view" "attack graph"]: 15 \\
    \hline
    NetSPA & \flushleft{Massachusetts Institute of Technology} & \flushleft{Not open source, difficult to obtain} & \hyperlink{MPAG}{MPAG} & O(NlgN) & Fair & 2006 & 493 & "Practical attack graph generation for network defense": 493 & ["NetSPA"]: 357 \\
    \hline
    SeaMonster & \flushleft{Norwegian Univ. of Science and Technology and SINTEF research foundation} & Open source & \hyperlink{AttackTree}{AT} & Polynomial & Fair & 2008 & 59 & "SeaMonster: Providing tool support for security modeling": 37 & ["seamonster" "attack tree"]: 59 \\
    \hline
    Cauldron & \flushleft{PROINFO Company, George Mason University} & Commercial Software & \hyperlink{EDG}{EDG} & O($N^3$) & Good & 2011 & 150 & \flushleft{"Cauldron mission-centric cyber situational awareness with defense in depth": 142} & [Cauldron "attack graph"]: 150 \\
    \hline
    FireMon & \flushleft{FireMon, Massachusetts Institute of Technology} & Commercial Software & \hyperlink{MPAG}{MPAG} & O(NlgN) & Good & 2012? & 16 & No paper & ["firemon" "attack graph"]: 16 \\
    \hline
    CySeMoL & \flushleft{Royal Institute of Technology, Stockholm, Sweden} & \flushleft{Not open source, difficult to obtain} & Unknown & Polynomial? & Not Provided & 2013 & 196 & \flushleft{"The Cyber Security modeling Language: A Tool for Assessing the Vulnerability of Enterprise System Architectures": 168} & ["cysemol"]: 196 \\
    \hline
    CyGraph & MITRE & \flushleft{Not open source, difficult to obtain} & Unknown & Scales well$^{\mathrm{(a)}}$ & Very Good$^{\mathrm{(b)}}$ & 2016 & 94 & "CyGraph: graph-based analytics and visualization for cyber security": 94 & ["cygraph" "attack graph"]: 46 \\
    \hline
    SAGE & \flushleft{Delft University of Technology, Netherlands, Rochester Institute of Technology, US} & Open source & Alert-driven & NA$^{\mathrm{(c)}}$ & Good & 2021 & 8 & \flushleft{"Alert-driven Attack Graph Generation using S-PDFA": 2} & ["SAGE" "attack graph"]: 8 \\
    \hline
    AttacKG & \flushleft{Zhejiang University, National University of Singapore, Northwestern University} & Open source & CTI-based & NA$^{\mathrm{(c)}}$ & Fair & 2021 & 0 & \flushleft{"Attackg: Constructing technique knowledge graph from cyber threat intelligence reports": 0} & ["AttacKG" "attack graph"]: 0 \\
    \hline
    \multicolumn{10}{l}{$^{\mathrm{(a)}}$Graph database complexity depends on the part of the graph traversed by the query, not the total number of nodes in the database}\\
    \multicolumn{10}{l}{$^{\mathrm{(b)}}$CyGraph includes graph dynamics, layering, grouping, filtering, and hierarchical views}\\
    \multicolumn{10}{l}{$^{\mathrm{(c)}}$Alert-driven and CTI-based AG generators don't refer to network size, therefore their scalability is irrelevant here}
    \end{tabular}
    \label{table:AttackGraphTools}
\end{table*}

\section{MulVAL Extensions}
\subsection{The MulVAL Framework} \label{MulvalFramework}
% Description of the Datalog syntax: Facts, and Rules.
MulVAL (multi-host, multi-stage vulnerability analysis language) is an open-source publicly available logic-based attack graph generation tool~\cite{ou2005logic}. 
MulVAL is based on the Datalog modeling language, which is a subset of the Prolog logic programming language. 
In MulVAL, Datalog is used to represent two types of entities:
\begin{itemize}
\item \textit{Facts}: network topology and configuration, security policy and known vulnerabilities
\item \textit{Rules}: also known as interaction rules, define the interactions between components in the network
\end{itemize}

Facts and rules are defined by applying a predicate $p$ to some arguments: $p(t_1,...,t_k)$.
Each $t_i$ can be either a constant or a variable.
Datalog syntax indicates that a constant is an identifier that starts with a lowercase letter, while a variable is one that starts with an uppercase letter.
A wildcard expression can be defined by the underscore character ('$\_$').
A sentence in MulVAL is defined as Horn clauses of literals:
$$L_0 :- L_1,...,L_n$$
    
$L_0$ is defined as the head, and $L_1,...,L_n$ are defined as the body of the sentence, respectively.
Each $L_i$ in the body can be either a fact or an interaction rule.
If the body $(L_1,...,L_n)$ literals are true, then the head ($L_0$) literal is also true.
A sentence with an empty body is called a fact.
For example, the following fact states that there is an identified vulnerability \texttt{CVE-2002-0392} in the \texttt{httpd} service running on \texttt{webServer01} instance:
\begin{lstlisting}[basicstyle=\ttfamily\footnotesize,language=Python]
vulExists(webServer01, "CVE-2002-0392", httpd).
\end{lstlisting}
A sentence with a nonempty body is called a rule.
For example, the rule in Listing~\ref{lst:fileprot} says that if a \texttt{User} has ownership of \texttt{Path} on \texttt{Host}, and if an owner of \texttt{Path} on \texttt{Host} has the specified \texttt{Access}, then the \texttt{User} on \texttt{Host} can have the specified \texttt{Access} to \texttt{Path}.
\begin{lstlisting}[basicstyle=\ttfamily\footnotesize,language=Python,label=lst:fileprot, caption=Interaction rule example]
localFileProtection(Host, User, Access, Path) :-
    fileOwner(Host, Path, User),
    ownerAccessible(Host, Access, Path).
\end{lstlisting}

% Review of LAG representation
\begin{figure*}[ht]
    \centering
    \includegraphics[width=0.95\textwidth]{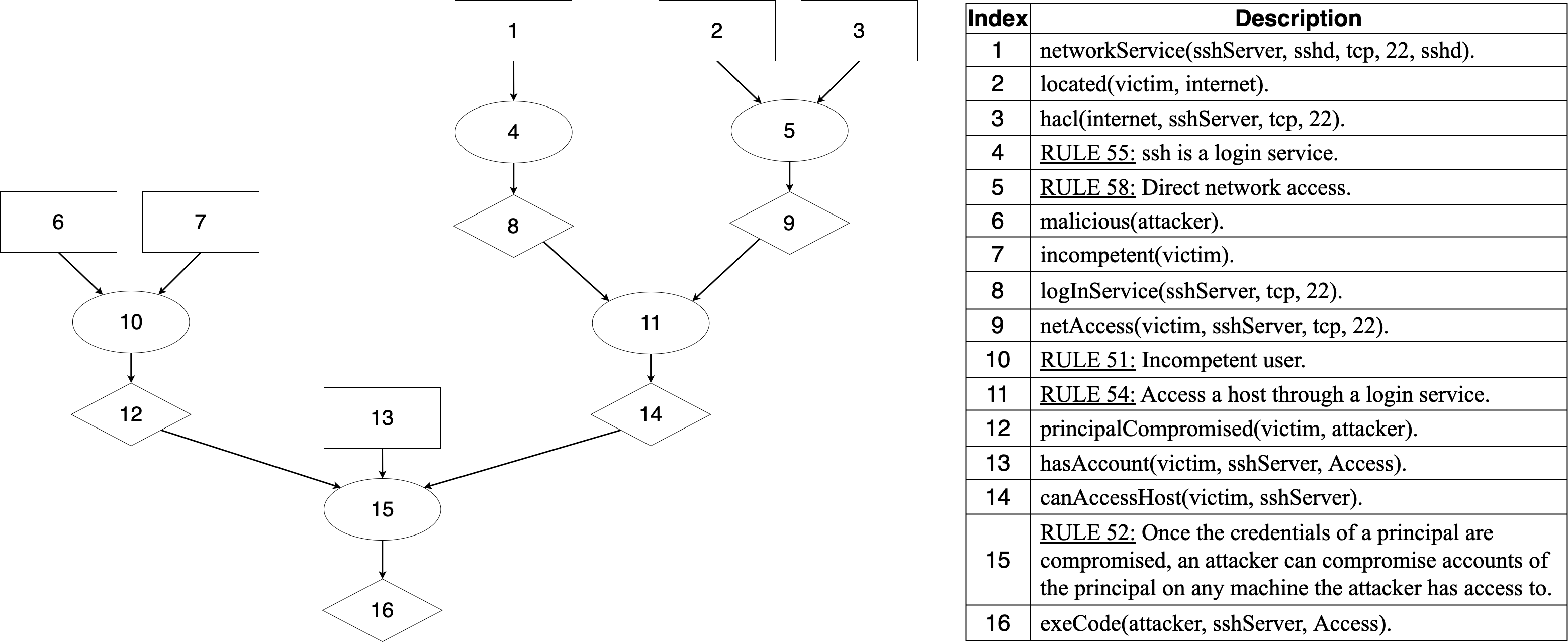}
    \caption{MulVAL example: code execution attack graph.}
    \label{fig:MulVALExample}
    %\Description{Attack graphs example with description of steps.}
\end{figure*}

Figure~\ref{fig:MulVALExample} presents an example of a LAG generated by MulVAL: a code execution attack via a remote service (sshd) performed by using a compromised user account.
In MulVAL the graph representation is constructed as follows:
\begin{itemize}
    \item Fact nodes (rectangles), also called primitive facts, represent the asset state, configuration, or network condition that must exist in order for the attack to exploit the vulnerability.
    \item Privilege nodes (diamonds), also called derived facts, represent the attack impact, e.g., the information or assets obtained by an attacker.
    \item Action nodes (circles), also called derivation or exploit nodes, represent the actions an attacker should perform to gain some privileges. 
\end{itemize}
To execute an exploit, which means performing some action, the attacker needs all of the privileges and facts that lead to that action.
As a result, an action node will lead to a single privilege node. 

\begin{figure}[ht]
    \centering
    \includegraphics[width=0.48\textwidth]{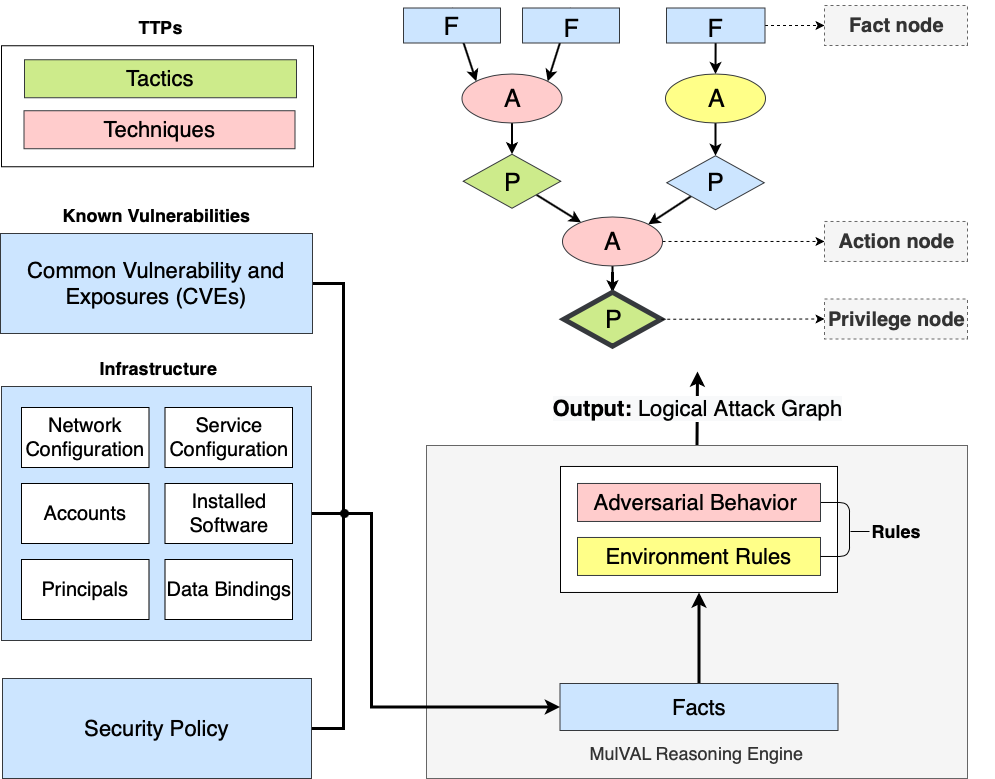}
    \caption{The MulVAL framework.}
    \label{fig:MulVALFramework}
    %\Description{MulVAL framework components.}
\end{figure}

% MulVAL required inputs
As depicted in Figure~\ref{fig:MulVALFramework}, MulVAL facts (which appear in blue) are constructed from:
\begin{itemize}
    \item Vulnerabilities
    \begin{itemize}
        \item Known vulnerabilities: CVEs registered in publicly available vulnerability databases, such as the NVD (National Vulnerability Database)~\cite{nvd}, VulDB (Vulnerability Database)~\cite{vuldb}, WhiteSource Vulnerability Database~\cite{whitesource}, etc.
        \item Unknown vulnerabilities: MulVAL facts can be used for simulating unknown vulnerabilities and testing network resilience against zero-day exploits.
        The following fact enables simulating unknown bug:
\begin{lstlisting}[basicstyle=\ttfamily\footnotesize,language=Python]
bugHyp(Host, Prog, ExploitRange, ExplConseq)
\end{lstlisting} 
    \end{itemize} 
    \item Infrastructure: the infrastructure setup, containing information regarding the current environment state, such as network configuration (e.g., network topology, firewall rules), service configuration, accounts, installed software, principals, and data bindings (symbolic names).
    \item Security policy: the security policy loaded into the reasoning engine.
\end{itemize}

The vulnerability and infrastructure configuration required can be collected using custom scripts or existing tools and services such as: Nessus~\cite{nessus} vulnerability scanner, host-based OVAL~\cite{baker2011oval} agents, etc.
The reasoning engine estimates the effect of the identified vulnerabilities on the system.
This estimation is performed by applying the defined set of interaction rules on the generated facts.
The MulVAL framework provides a default set of various interaction rules~\cite{mulval}.
These rules are represented as action nodes in the LAG and can be categorized into two types:
\begin{itemize}
    \item Environment rules (in yellow): describing additional security-related facts (see Definition~\ref{def:envrule} in Section~\ref{definitions}). For example, index 4 in Figure~\ref{fig:MulVALExample} identifies \texttt{sshd} as a login service.
    \item Adversarial behavior (in red): describing an attack technique. For example, index 15 in Figure~\ref{fig:MulVALExample} enables the attacker to apply a code execution technique.
\end{itemize}

Unlike CVEs, principals, network configuration, etc., the set of rules defining the behavior of the adversary and the mechanics of the environment rarely change. 
Rules can be extended to represent known tactics, techniques, and procedures (TTPs).
However, procedures are a highly detailed description of a technique and consequently are rarely modeled in the LAG.
In addition, rules can be used to represent different IT advances such as near-field communication or cloud technologies~\cite{stan2020extending, mensah2019generation, albanese2017computer}.
Tactics (which appear in green in Figure~\ref{fig:MulVALFramework}) describe the short-term goals of the attacker.
They are represented as privilege nodes that are created by an adversarial behavior.
Each of these nodes advances the attacker towards the final goal which is achieved in the final privilege node.
Techniques (in red) describe the attacker's actions.

Being generic and extensible, LAGs support a large class of threat models, characterized by attacker's goals, capabilities, and resources.  
Attackers may have arbitrary goals represented as assets~\cite{ou2005logic}. 
LAGs may also support the various levels of attacker capabilities if they are defined as preconditions for exploits~\cite{malzahn2020automated, wang2020cvss}.  
However, due to the common delete-free relaxation in LAG solvers, the modeling of attacker resources may be challenging. 

% The XSB system and its advantages.
MulVAL uses the XSB (Extended Stony Brook) environment~\cite{rao1997xsb}, which supports a declarative style of logic programming of Datalog programs, called table execution. 
XSB enables effective dynamic programming that avoids recomputation of previously calculated facts, thus enabling the reasoning engine to scale well with the size of the network.

% MulVAL advantages over other AG generation tools
In 2013, Yi et al.~\cite{yi2013overview} compared several academic and commercial attack graph generation tools (TVA, Attack Graph Toolkit, NetSPA, MulVAL, Cauldron, FireMon, and Skybox View).
The authors concluded that MulVAL is the most extendable and scalable framework; commercial tools may be more scalable and user-friendly, however they are not open-source and are thus less suitable for academic research.
In our review we add five additional attack graph generation tools to the comparison.
Table~\ref{table:AttackGraphTools} is based on the comparison made by Yi et al.~\cite{yi2013overview}, with the addition of SeaMonster, CySeMoL, CyGraph, SAGE, and AttacKG, and four additional columns: Year (the year in which the tool was first published), Number of References (the larger value of the next two columns), Paper Search (the number of Google Scholar citations for the tool's main paper between 2005-2021), and Tool Search (the results of a search of the tool's keyword(s) in Google Scholar between 2005-2021).

Table~\ref{table:AttackGraphTools} shows that MulVAL has several advantages:
\begin{itemize}
    \item Availability: it is open-source.
    \item Scalability: its execution time is $O(n^2)$ relative to the size of the network~\cite{ou2006scalable}.
    \item Extensibility: its underlying reasoning engine is written in a logical programming language, which enables users to extend functionality by writing custom rules.
    \item Compatibility: it leverages public vulnerability resources which are continuously updated.
    \item Broad acceptance: as depicted in Table~\ref{table:AttackGraphTools}, MulVAL is the tool most referred to by researchers.
\end{itemize}
Therefore, in this study, we focus on the MulVAL attack graph framework and its reasoning engine in particular. 

\subsection{Definitions} \label{definitions}
Our first goal in this paper is to conduct a thorough survey and identify all papers that extend MulVAL and add new interaction rules to describe new attacks.
We begin by providing some formal definitions, using the ``Exploitation for Privilege Escalation" MITRE ATT\&CK Technique.
Each Technique can be implemented using one or more attack procedures.
For example, the exploitation for privilege escalation technique can be implemented by executing code on a host where software with a vulnerability exists or by injecting a command into a host with a bad configuration.

The first procedure can be represented by MulVAL rules as presented in Listing~\ref{lst:execcode}:
\begin{lstlisting}[basicstyle=\ttfamily\footnotesize,language=Python,label=lst:execcode, caption=A technique procedure expression]
execCode(Prin, Host, Perm) :-
    malicious(Prin),
    execCode(Prin, Host, Perm2),
    vulExists(Host, Software, localExploit, privEsc),
    setuidProgram(Host, Software, Perm).
\end{lstlisting}

The description of a ATT\&CK Technique implies one or more attack procedures. 
Such procedures may include interactions between multiple entities, such as users or computer resources.
If a set of interaction rules encodes all relevant interactions to describe an attack procedure implied by a Technique, we say that this set of rules expresses the Technique.

\begin{definition}[Expressing a technique]
\label{def:express}
A set of MulVAL interaction rules $SIR = \{R_1, R_2, ..., R_n\}$ \textbf{expresses} a MITRE ATT\&CK Technique if SIR is a minimal set that is sufficient to represent a Technique's procedure in an attack graph.
The reason for minimal set is efficiency and clarity.
When there is a SIR expressing a Technique, MulVAL \textbf{covers} this Technique.
\end{definition}

For example, according to the description of the Exploitation for Privilege Escalation Technique, ``Adversaries may exploit software vulnerabilities in an attempt to collect elevate privileges.
Exploitation of a software vulnerability occurs when an adversary takes advantage of a programming error in a program, service, or within the operating system software or kernel itself to execute adversary-controlled code."
Listing~\ref{lst:execcode} \textbf{expresses} a procedure of this Technique.
Since each Technique may have different procedures, it can be expressed with different sets of interaction rules (SIRs).

\begin{definition}[Partial expression]
If the number of rules in the expressing set $|SIR| > 1$, any subgroup $SIR' \subset SIR$ \textbf{partially expresses} the Technique.
\end{definition}

In the above example, the following rule \textbf{partially expresses} the Technique:
\begin{lstlisting}[basicstyle=\ttfamily\footnotesize,language=Python]
vulExists(Host, Software, localExploit, privEsc)
\end{lstlisting}

It should be mentioned that the same interaction rule can be used in different sets (SIRs) and partially express different Techniques.

\begin{definition}[Environment rule] \label{def:envrule}
An \textbf{environment rule} is a predicate describing a security-related configuration, a formal software vulnerability, or a security policy defined by system administrators. Environment rules are used as input to MulVAL.
An environment rule can be a primitive predicate, which will be referred to as a \textbf{primitive environment rule} (or simply a \textbf{fact}) or a derived predicate, which will be referred as a \textbf{derived environment rule}.
\end{definition}

For example, the following predicate is a \textbf{primitive environment rule (fact)} describing that a service \texttt{Prog} is running on \texttt{Host} as \texttt{User} and listening on \texttt{Port} of \texttt{Protocol}:
\begin{lstlisting}[basicstyle=\ttfamily\footnotesize,language=Python]
networkService(Host, Prog, Protocol, Port, User)
\end{lstlisting}

Listing~\ref{lst:vulexists} is a \textbf{derived environment rule} describing that if a \texttt{Prog} running on \texttt{Host} depends on \texttt{Library}, which has a vulnerability, then the \texttt{Prog} has the same vulnerability.
\begin{lstlisting}[basicstyle=\ttfamily\footnotesize,language=Python,label=lst:vulexists, caption=Derived environment rule example]
vulExists(Host, Prog, Consequence) :- 
    vulExists(Host, Library, Consequence),
    dependsOn(Host, Prog, Library).
\end{lstlisting}

\begin{definition}[Building block] A derived predicate is called a \textbf{building block} if it is a general attack step that can be used in many SIRs, i.e., it can partially express many Techniques.
\end{definition}

Listing~\ref{lst:canaccesshost} is a \textbf{building block} describing that if a user \texttt{Prin} has access to a \texttt{Host} from any source computer (\texttt{\_Src}) on a \texttt{Port} of \texttt{Protocol}, and the \texttt{Host} enables login service in the same port and protocol, then user \texttt{Prin} can log in to the \texttt{Host}.
\begin{lstlisting}[basicstyle=\ttfamily\footnotesize,language=Python,label=lst:canaccesshost, caption=Building block example]
canLogInHost(Prin, Host) :-
    logInService(Host, Protocol, Port),
    netAccess(Prin, _Src, Host, Protocol, Port).
\end{lstlisting}

This is a building block, since accessing a host can be a step in many attack procedures (SIRs).

\subsection{Search Method} \label{SearchMethod}
\begin{figure}[ht]
    \centering
    \begin{subfigure}[b]{\columnwidth}
    \centering
    \includegraphics[width=0.95\columnwidth]{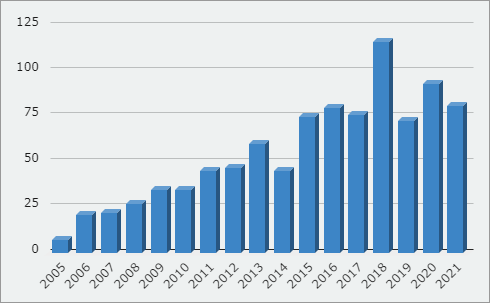}
    \caption{Number of papers mentioning MulVAL between 2005-2021 (total: 938).}
    \label{fig:MentioningMulVAL}
    \end{subfigure}
    \centering
    \begin{subfigure}[b]{\columnwidth}
    \centering
    \includegraphics[width=0.95\columnwidth]{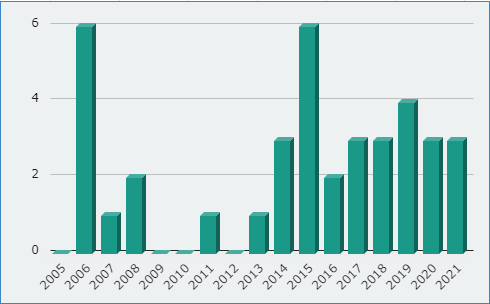}
    \caption{Number of papers extending MulVAL between 2005-2021 (total: 38).}
    \label{fig:ExtendingMulVAL}
    \end{subfigure}
    \caption{A timeline of MulVAL publications.}
    %\Description{MulVAL mentioning and extending number of papers.}
\end{figure}

To find all of the MulVAL extensions, we performed a systematic literature review.
The goal of our review was to identify all of the academic papers that present MulVAL extensions.
Since the papers do not always explicitly mention the fact that they are extending MulVAL, we searched Google Scholar for a single phrase ``MulVAL," excluding patents and quotes.
Since the original paper describing MulVAL was published in 2005, the search was limited to the years 2005-2021.
We found 938 papers available online on February 2022.
The next step was to identify and remove all of the papers that mentioned MulVAL but did not add any new interaction rules.
By manually examining these papers, we identified a set of 38 papers in which the authors presented an extension to the MulVAL framework either by introducing new interaction rules or by describing a method for defining new interaction rules.
Figure~\ref{fig:MentioningMulVAL} shows the number of papers mentioning MulVAL each year, from 2005 to 2021, and Figure~\ref{fig:ExtendingMulVAL} shows the number of papers extending MulVAL during that time.

\subsection{Extension Findings} \label{ExtensionFindings}
From the set of 38 papers that presented an extension to the MulVAL framework, 21 papers defined new interaction rules representing new attack procedures.
Together, the base MulVAL paper and these papers defined a total of 349 predicates: 92 environment rules (describing security-related configuration information, formal vulnerabilities, or security policies defined by system administrators) and 257 interaction rules defining new attack procedures.
Table~\ref{table:MulVALExtension} presents a list of all of the papers extending MulVAL, including the extension's field, the methodology used, the number of times the paper has been cited, and the number of environment and interaction rules.

\begin{table*}[htb]
    \centering
    \caption{Papers extending MulVAL}
    \scriptsize
    \begin{tabular}{|| m{5em} | m{5em} | m{12em} | m{11em} | m{6em} | m{5em} | m{5em} | m{5em} ||} 
    \hline
    \textbf{Paper} & \textbf{Year} & \textbf{Venue} & \textbf{Extension Field} & \textbf{Methodology} & \textbf{No. of Citations} & \textbf{No. of Env. Rules} & \textbf{No. of SIRs} \\ [0.5ex] 
    \hline \hline
    \cite{ou2005logic} & 2005 & Princeton University & NA (base MulVAL paper) & NA & 61 & 35 & 26 \\ 
    \hline
    \cite{ou2006scalable} & 2006 & ACM Conference & Framework & NA & 757 & 0 & 0 \\
    \hline
    \cite{bacic2006mulval} & 2006 & CINNABAR Networks & Framework & NA & 13 & 0 & 0 \\
    \hline
    \cite{bhatt2006model} & 2006 & NATO Security, IOS & Access Rules for Apache & Ad Hoc & 1 & 0 & 3 \\
    \hline
    \cite{govindavajhala2006status} & 2006 & Princeton University & Access Rules for Windows & Ad Hoc & 3 & 8 & 1 \\
    \hline
    \cite{govindavajhala2006windows} & 2006 & Princeton University & Access Rules for Windows & Ad Hoc & 39 & 0 & 4 \\
    \hline
    \cite{govindavajhala2006formal,govindavajhala2007automatic} & 2006, 2007 & Princeton University & Access Rules for Windows & Ad Hoc & 4 & 18 & 8 \\
    \hline
    \cite{homer2008attack} & 2008 & Kansas State University & Enterprise & Ad Hoc & 12 & 1 & 2 \\
    \hline
    \cite{saha2008extending} & 2008 & ACM Conference & Framework & NA & 58 & 0 & 0 \\
    \hline
    \cite{ou2011quantitative} & 2011 & Springer & Enterprise & Ad Hoc & 52 & 0 & 2 \\
    \hline
    \cite{almohri2013high,almohri2015security} & 2013, 2015 & IEEE & Mobile Connectivity & Ad Hoc & 51 & 0 & 1 \\
    \hline
    \cite{liu2014model,liu2014relating,liu2015logic,liu2015probabilistic} & 2014, 2015 & IEEE & Framework & NA & 37 & 0 & 0 \\
    \hline
    \cite{sun2014inferring,sun2018probabilistic} & 2014, 2018 & Springer & Cloud Computing & Methodical & 9 & 6 & 4 \\
    \hline
    \cite{sembiring2015network} & 2015 & Electrical Eng. \& Informatics & Framework & NA & 19 & 0 & 0 \\
    \hline
    \cite{jilcott2015securing} & 2015 & IEEE & Enterprise & Ad Hoc & 4 & 1 & 1 \\
    \hline
    \cite{dongestablishing,dong2016right} & 2015, 2016 & Springer & Infected USB & Ad Hoc & 0 & 0 & 1 \\
    \hline
    \cite{acosta2016augmenting} & 2016 & IEEE & Enterprise & Ad Hoc & 11 & 2 & 3 \\
    \hline
    \cite{jing2017augmenting} & 2017 & IEEE & Rule Generation & Methodical & 7 & 0 & 0 \\
    \hline
    \cite{sun2017towards} & 2017 & Springer & Cloud Computing & Methodical & 13 & 0 & 2 \\
    \hline
    \cite{albanese2017computer} & 2017 & Springer & Cloud Computing & Methodical & 6 & 0 & 0 \\
    \hline
    \cite{anderson2018determining} & 2018 & Eastern Washington Univ. & Framework & NA & 0 & 0 & 0 \\
    \hline
    \cite{cao2018assessing} & 2018 & Springer & Graph Connectivity & NA & 11 & 0 & 0 \\
    \hline
    \cite{mensah2019generation} & 2019 & CentraleSup{\'e}lec & Cloud Computing & Methodical & 1 & 8 & 5 \\
    \hline
    \cite{appana2019first} & 2019 & IEEE & Framework & NA & 0 & 0 & 0 \\
    \hline
    \cite{khakpour2019towards} & 2019 & IEEE & Enterprise & Ad Hoc & 8 & 0 & 8 \\
    \hline
    \cite{inokuchi2019design} & 2019 & ACM Conference & Enterprise & Methodical & 8 & 1 & 6 \\
    \hline
    \cite{stan2020extending} & 2020 & IEEE & Enterprise & Methodical & 12 & 0 & 60 \\
    \hline
    \cite{zhou2020security} & 2020 & DiVA & Data Criticality & Methodical & 0 & 3 & 11 \\
    \hline
    \cite{mccormack2020security} & 2020 & IEEE & 3D Printer & Ad Hoc & 2 & 10 & 54 \\
    \hline
    \cite{stan2021heuristic} & 2021 & IEEE & Framework & Methodical & 1 & 0 & 1 \\
    \hline
    \cite{binyamini2021framework} & 2021 & ACM Conference & Rule Generation & Methodical & 0 & 0 & 0 \\
    \hline
    \cite{bitton2021evaluating} & 2021 & arXiv Preprint & Machine Learning & Methodical & 0 & 0 & 53 \\
    \hline \hline
    Total &  &  &  &  &  & 92 & 257 \\
    \hline
    \end{tabular}
    \label{table:MulVALExtension}
\end{table*}

\noindent\textbf{Framework improvements:}
Ou et al.~\cite{ou2006scalable} demonstrated how to produce a derivation trace in the MulVAL logic-programming engine and how the trace can be used to generate a LAG in quadratic time.
Bacic et al.~\cite{bacic2006mulval} extended MulVAL to improve network representation and derive rules in a more intuitive manner.
Saha~\cite{saha2008extending} extended the MulVAL framework to include complex security policies and extended the LAG concept to include justification as to why a negated sub-goal failed.
Liu et al.~\cite{liu2014model, liu2014relating, liu2015logic, liu2015probabilistic} used evidence obtained from security events to construct an attack scenario and build an evidence graph.
Sembiring et al.~\cite{sembiring2015network} introduced three methods to improve the MulVAL framework: employing the Common Vulnerability Scoring System (CVSS) to calculate the probability of vulnerability variables and the Common Configuration Scoring System (CCSS) to calculate the probability of system security configuration vulnerabilities; introducing the concept of interdependence between vulnerability variables in Bayesian senses; and analyzing the impact of a change in the system security configuration on the probability of vulnerabilities in the context of Bayesian probability.
Anderson~\cite{anderson2018determining} explored enhancing estimations of factor analysis of information risk vulnerability by modeling interactions between threat actors and assets through attack graphs.
Appana et al.~\cite{appana2019first} proposed applying a ranking algorithm on the mission impact graph, based on the MulVAL attack graph.
Stan et al.~\cite{stan2021heuristic} proposed a method that expresses the risk of the system, using an extended attack graph model that considers the prerequisites and consequences of exploiting a vulnerability, examines the attacker’s potential lateral movements, and expresses the physical network topology as well as vulnerabilities in network protocols.

\noindent\textbf{Access rules:}
Bhatt et al.~\cite{bhatt2006model} presented a model-driven technique for automated policy-based access analysis and added three access rules for Apache.
Govindavajhala et al.~\cite{govindavajhala2006status, govindavajhala2006windows, govindavajhala2006formal, govindavajhala2007automatic} suggested separating scanning from analysis to reduce the size of code running in privileged mode.
They also demonstrated how to extend the MulVAL framework to reason about the security of a network with hosts running disparate operating systems.
In particular, they illustrated 39 reasoning rules for Windows to find misconfigurations of the access control lists.

\noindent\textbf{Enterprise:}
Homer et al.~\cite{homer2008attack} presented methodologies that can automatically identify and trim portions of an attack graph that do not help a user understand the core security problems.
Ou et al.~\cite{ou2011quantitative} presented an approach which, given component metrics that characterize the likelihood that individual vulnerabilities can be successfully exploited, computes a numeric value representing the cumulative likelihood for an attacker to succeed in gaining a specific privilege or carrying out an attack in the network.
Jilcott~\cite{jilcott2015securing} presented a technology that automatically maps and explores the firmware/software architecture of a commodity IT device and then generates attack scenarios for the device.
Acosta et al.~\cite{acosta2016augmenting} augmented MulVAL to incorporate network layer misconfigurations.
In particular, they presented ARP spoofing and route hijacking scenarios.
Khakpour et al.~\cite{khakpour2019towards} defined several rules for the exploitation and propagation of vulnerabilities.
Inokuchi et al.~\cite{inokuchi2019design} proposed a methodical procedure for defining new interaction rules, and they applied the method to define four categories of behavior: privilege escalation, credential access, lateral movement, and execution.
Stan et al.~\cite{stan2020extending} presented an extended network security model for MulVAL that considers the physical network topology, supports short-range communication protocols, models vulnerabilities in the design of network protocols, and models specific industrial communication architectures.
They also introduced an extensive list of 60 new interaction rules.

\noindent\textbf{Cloud:}
Sun et al.~\cite{sun2014inferring, sun2018probabilistic} referred to two cloud risks: virtual machine (VM) images may be shared between different users, and VMs owned by different tenants may co-reside on the same physical host machine.
Sun et al.~\cite{sun2017towards} dealt with the gap between mission impact assessment and cyber resilience in the context of cloud computing.
The authors bridged this gap by developing a graphical model that interconnects the mission dependency graphs and cloud-level attack graphs.
Albanese et al.~\cite{albanese2017computer} proposed building cross-layer Bayesian networks to infer the stealthy bridges between enterprise network islands in clouds.
Mensah~\cite{mensah2019generation} extended MulVAL to include cloud virtualization vulnerabilities.

\noindent\textbf{Rule generation:}
Jing et al.~\cite{jing2017augmenting} presented a tool that can parse vulnerability descriptions, as provided in the CVE, to retrieve relevant information for generating interaction rules that can be incorporated into MulVAL.
Binyamini et al.~\cite{binyamini2021framework} presented an automated framework for modeling new attack techniques from the textual description of a security vulnerability.
Their framework enables the automatic generation of MulVAL interaction rules from the NVD.

\noindent\textbf{Other:}
Almohri et al.~\cite{almohri2013high, almohri2015security} addressed the problem of statically performing a rigorous assessment of a set of network security defense strategies, with the goal of reducing the probability of a successful large-scale attack in a complex, dynamically changing network architecture.
Dong et al.~\cite{dongestablishing, dong2016right} presented common input scenarios for different model-based security assessment tools.
Cao et al.~\cite{cao2018assessing} proposed a business process impact assessment method, which measures the impact of an attack targeting a business-process-support enterprise network.
Zhou~\cite{zhou2020security} extended the security risk analysis with data criticality and introduced 14 new interaction rules.
McCormack et al.~\cite{mccormack2020security} focused on identifying security threats to networked 3D printers.
Bitton el al.~\cite{bitton2021evaluating} extended MulVAL with 54 interaction rules to model attacks on machine learning production systems.

\section{Coverage of Attack Scenarios in MulVAL}

To estimate the coverage of attack scenarios by MulVAL, we decided to map MulVAL interaction rules to MITRE ATT\&CK.
Section~\ref{MitreAttack} describes MITRE ATT\&CK, and Section~\ref{TechniquesCovered} presents the expressed ATT\&CK Techniques in MulVAL.

\subsection{MITRE ATT\&CK} \label{MitreAttack}
The Mitre Corporation (MITRE) is an American nonprofit organization dedicated to bringing innovative ideas into existence in different areas related to safety and security~\cite{mitrewebsite}.
MITRE ATT\&CK (Adversarial Tactics, Techniques, and Common Knowledge) is a knowledge base of adversarial tactics and techniques based on real-world observations~\cite{strom2018mitre}.
The ATT\&CK knowledge base is used as a foundation for the development of specific threat models and methodologies in the cyber security community.
ATT\&CK provides a common taxonomy for both offense and defense, and has become a useful conceptual tool across many cyber security disciplines, helping improve network and system defenses against intrusions.

MITRE ATT\&CK reflects the various phases of an adversary’s attack lifecycle and focuses on how external adversaries compromise and operate within computer information networks.
ATT\&CK is a behavioral model that consists of the following core components:
\begin{itemize}
    \item Tactics, denoting short-term, tactical adversarial goals during an attack
    \item Techniques, describing the means by which adversaries achieve tactical goals
    \item Sub-techniques, describing more specific means by which adversaries achieve tactical goals at a lower level than techniques
    \item Documented adversary use of techniques, their procedures, and other metadata
\end{itemize}

ATT\&CK has different use cases, including: adversary emulation, red teaming, behavioral analytic development, defensive gap assessment, SOC (security operations center) maturity assessment, and cyber threat intelligence (CTI) enrichment.
For example, Oosthoek et al.~\cite{oosthoek2019sok} plotted a sample of 951 Windows malware families on the ATT\&CK framework to obtain insights on trends in attack techniques used to target Windows.
Maynard et al.~\cite{maynard2020big} created an ATT\&CK model of a hacktivist and mapped the threat to critical infrastructure in order to better define the skills and methods a hacker might employ.
To assist developers and administrators in cultivating an attacker mindset, Munaiah et al.~\cite{munaiah2019characterizing} used the MITRE ATT\&CK framework to characterize an attacker's campaign in terms of Tactics and Techniques.
Analysts can use the ATT\&CK framework to structure intelligence about adversary behavior, and defenders can structure information about what behavior they can detect and mitigate~\cite{nickels2018using}.
By overlaying information from several adversary groups, they can create threat-based awareness of the gaps adversaries exploit.
Such analysis also improves CTI actionability for decision makers.

Beside MITRE ATT\&CK, there are other known methods of threat modeling.
Shevchenko~\cite{shevchenko2018threat} summarized 12 threat-modeling methods, including Microsoft STRIDE (spoofing, tampering, repudiation, information disclosure, denial of service, elevation of privilege), PASTA (process for attack simulation and threat analysis), and LINDDUN (linkability, identifiability, non-repudiation, detectability, disclosure of information, unawareness, non-compliance).
Another threat model is the cyber kill chain, a traditional model used to analyze cyber security threats, malware infecting of computer systems, covert and illegitimate channels found on a network, and insider threats~\cite{khan2018cognitive}.
These models are useful for gaining increased understanding of high-level processes and adversary goals.
However, the MITRE ATT\&CK model is more effective at conveying the individual actions performed by adversaries, how one action relates to another and to tactical adversarial objectives, and how the actions correlate with data sources, defenses, and configurations.

\begin{figure}[ht]
    \centering
    \includegraphics[width=0.48\textwidth]{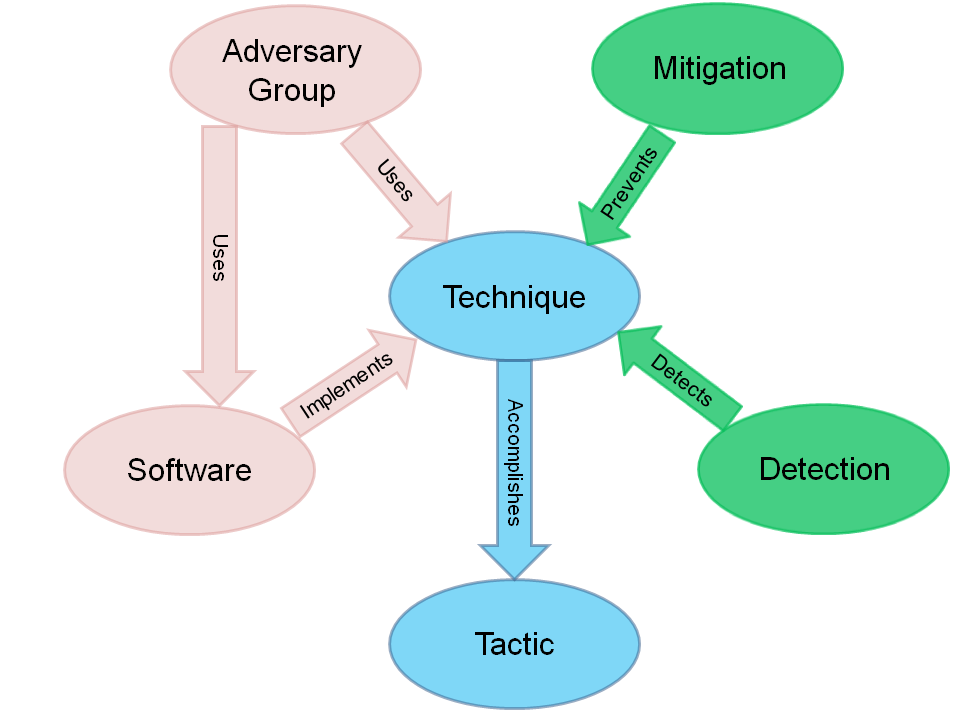}
    \caption{Relevant relationships between MITRE ATT\&CK concepts.}
    \label{fig:AttackModelRelationships}
\end{figure}

Representing an attack in terms of Tactics, Techniques, and Sub-techniques provides a means of balancing the technical details in the Technique description and the attack goals represented by the Tactics.
Tactics represent the “why” of an ATT\&CK Technique or Sub-technique - the adversary’s tactical objective, i.e., the desired outcome of performing an action.
Techniques represent the “how” - the actions an adversary performs to achieve a tactical objective.
Sub-techniques further break down behaviors described by Techniques. 
Procedures are the specific implementations that adversaries use to apply Techniques or Sub-techniques.
In addition to textual descriptions, metadata, Sub-techniques, and Procedures, a Technique may also include:
\begin{itemize}
    \item Group - known groups of adversaries that are tracked and reported on in threat intelligence reports.
    \item Software - tools and malware used by adversaries.
    \item Mitigation - security concepts and classes of technologies that can be used to prevent a technique from being successfully executed.
    \item Detection - methods for detecting  the use of a Technique by an adversary. 
\end{itemize}
The relationships between these concepts are depicted in Figure~\ref{fig:AttackModelRelationships}.
Adversary Group and Software (on the left) are related to the attacker, while Detection and Mitigation (on the right) are related to the defender.

ATT\&CK is organized in a series of technology domains -- the ecosystem an adversary operates within.
To date, MITRE has defined three technology domains: Enterprise, Mobile, and ICS (industrial control system).
In this work, we focus on the Enterprise domain.
In the major version of MITRE ATT\&CK Enterprise from October 2020 (ATT\&CK Content version 8.0), two pre-attack Tactics were added, bringing the total number of Tactics to 14~\cite{mitreattackwebsite}.
Regarding ATT\&CK's coverage, it is important to note that in general, coverage of every ATT\&CK Technique is unrealistic~\cite{nickels2019savvy}.
Similarly, since each ATT\&CK Technique may have many implementation procedures that can be used by an adversary, and we cannot possibly know all of them, it is unrealistic to cover all procedures for a given technique.

\subsection{Expressed ATT\&CK Techniques in MulVAL} \label{TechniquesCovered}
Our second goal in this paper is to map MulVAL rules to MITRE ATT\&CK Techniques.
Mapping between the most commonly-used attack graph generation tool (i.e., MulVAL) and the MITRE ATT\&CK threat model will enable researchers and security administrators to handle additional realistic attack scenarios.

During the mapping process, we encountered an issue with Techniques that are associated with more than one Tactic.
These Techniques are presented in Table~\ref{table:TechniquesInMultiTactics}.
When implementing interaction rules to express a Technique, in some cases, the implementation can change depending on the Tactic (which is the attack goal).
Since each Technique in ATT\&CK is supposed to be unique, in cases in which the Tactic may affect the Technique, it seems that it may be better to define a different Technique for each Tactic.
The last column in Table~\ref{table:TechniquesInMultiTactics} reflects this issue, indicating whether a different Technique for each Tactic is recommended.

\begin{table*}[h!tb]
    \centering
    \caption{MITRE ATT\&CK Techniques listed under two or more Tactics}
    \scriptsize
    \begin{tabular}{| c | c | c |} 
    \hline
    \textbf{Technique} & \textbf{Tactics} & \textbf{Separation Recommended}\\ [0.8ex] 
    \hline
    Abuse Elevation Control Mechanism & Privilege Escalation, Defense Evasion & No \\ 
    \hline
    Access Token Manipulation & Privilege Escalation, Defense Evasion & No \\
    \hline
    BITS Jobs & Persistence, Defense Evasion & No \\
    \hline
    Boot or Logon Autostart Execution & Persistence, Privilege Escalation & Yes \\
    \hline
    Boot or Logon Initialization Scripts & Persistence, Privilege Escalation & Yes \\
    \hline
    Create or Modify System Process & Persistence, Privilege Escalation & Yes \\
    \hline
    Event Triggered Execution & Persistence, Privilege Escalation & Yes \\
    \hline
    External Remote Services & Initial Access, Persistence & No \\
    \hline
    Group Policy Modification & Privilege Escalation, Defense Evasion & No \\
    \hline
    Hijack Execution Flow & Persistence, Privilege Escalation, Defense Evasion & No \\
    \hline
    Input Capture & Credential Access, Collection & No \\
    \hline
    Man-in-the-Middle & Credential Access, Collection & No \\
    \hline
    Modify Authentication Process & Defense Evasion, Credential Access & No \\
    \hline
    Network Sniffing & Credential Access, Discovery & Yes \\
    \hline
    Pre-OS Boot & Persistence, Defense Evasion & No \\
    \hline
    Process Injection & Privilege Escalation, Defense Evasion & No \\
    \hline
    Replication Through Removable Media & Initial Access, Lateral Movement & Yes \\
    \hline
    Scheduled Task/Job & Execution, Persistence, Privilege Escalation & Yes \\
    \hline
    Software Deployment Tools & Execution, Lateral Movement & No \\
    \hline
    Traffic Signaling & Persistence, Defense Evasion, Command and Control & No \\
    \hline
    Use Alternate Authentication Material & Defense Evasion, Lateral Movement & No \\
    \hline
    Valid Accounts & Initial Access, Persistence, Privilege Escalation, Defense Evasion & Yes \\
    \hline
    Virtualization/Sandbox Evasion & Defense Evasion, Discovery & No \\
    \hline
    \end{tabular}
    \label{table:TechniquesInMultiTactics}
\end{table*}

In addition, as mentioned above, McCormack et al.~\cite{mccormack2020security} defined a list of 64 new interaction rules, focusing on identifying security threats to networked 3D printers, and Bitton el al.~\cite{bitton2021evaluating} defined a list of 54 interaction rules to model adversarial machine learning.
Attacks on 3D printing and machine learning are not yet covered in ATT\&CK Techniques.
 
\begin{figure}[ht]
    \centering
    \includegraphics[width=0.48\textwidth]{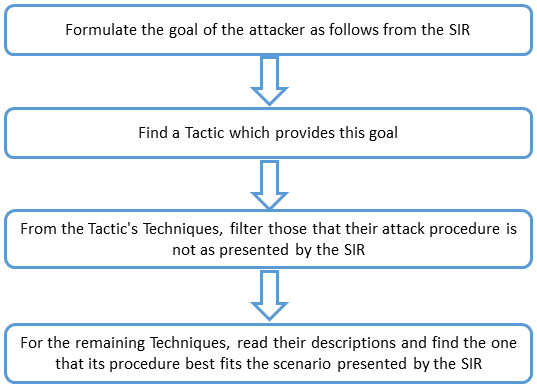}
    \caption{Mapping a SIR to a Technique.}
    \label{fig:MapSIRtoTechnique}
\end{figure}

To map MulVAL sets of interaction rules (SIRs) to ATT\&CK Techniques, we manually analyze each SIR, and according to the SIR's description and predicates, we first map it to a Tactic, and then to a specific Technique.
Figure~\ref{fig:MapSIRtoTechnique} presents the method of mapping a SIR to a Technique.
If a SIR is a general rule that may partially express many techniques, we call it a building block and do not connect it to a particular Technique.
In some cases, the same SIR can be used to express different Techniques.
In these cases, we connect the SIR to all of those Techniques.
Since the SIRs were defined in different studies for different purposes, there are some Techniques with a few SIRs, each expressing a different procedure, and many Techniques remain uncovered.
Table~\ref{table:ExpressedTechniques} presents the number of Techniques in each Tactic and the number of expressed Techniques in MulVAL for each Tactic.

\begin{table}[htbp]
    \centering
    \caption{Expressed Techniques in Tactics}
    \scriptsize
    %\small
    \begin{tabular}{| c | c | c |} 
    \hline
    \textbf{Tactic} & \textbf{No. of Techniques} & \textbf{No. of Expressed Tech.} \\
    \hline
    Reconnaissance & 10 & 0 \\ 
    \hline
    Resource Development & 6 & 0 \\
    \hline
    Initial Access & 9 & 5 \\
    \hline
    Execution & 10 & 5 \\
    \hline
    Persistence & 18 & 2 \\
    \hline
    Privilege Escalation & 12 & 4 \\
    \hline
    Defense Evasion & 37 & 3 \\
    \hline
    Credential Access & 14 & 7 \\
    \hline
    Discovery & 25 & 3 \\
    \hline
    Lateral Movement & 9 & 3 \\
    \hline
    Collection & 17 & 3 \\
    \hline
    Command and Control & 16 & 0 \\
    \hline
    Exfiltration & 9 & 3 \\
    \hline
    Impact & 13 & 4 \\
    \hline\hline
    Total & 205 & 42 \\
    \hline
    \end{tabular}
    \label{table:ExpressedTechniques}
\end{table}

There are some Tactics that are covered more, e.g., Initial Access, Execution, and Credential Access, and there are some Tactics that are covered less or are not covered at all, e.g., Reconnaissance, Resource Development, and Command and Control.
We provide the complete list of MulVAL rules and their mappings to MITRE ATT\&CK Enterprise Techniques\footnote{The list of MITRE ATT\&CK Enterprise Techniques and the MulVAL rules mapped to each Technique is available at: \url{https://github.com/dtayouri/MulVAL-MITRE/blob/main/ATT\%26CK\%20Enterprise\%20Techniques\%20with\%20MulVAL\%20IR.xlsx}}.
To generate attack graphs representing specific attack scenarios, one can only use the interaction rules mapped to the relevant Tactics or Techniques.
For example, to assess just the risks of initial access scenarios in a network, one should use the interaction rules mapped to the Initial Access Tactic's Techniques.
Mapping all of the MulVAL rules to ATT\&CK Techniques also enables actionable insights: as mentioned in the previous section, Techniques' Detection and Mitigation can be used to detect and mitigate the risks found with MulVAL rules that were part of the attack graph generation.

Table \ref{table:ExpressedTechniquesAnalysis} presents the list of Enterprise Techniques expressed with MulVAL rules, the number of SIRs mapped to them, and popularity analysis of expressed Techniques, as described below.
There are Techniques with many expressed procedures, e.g., Man-in-the-Middle, Exploitation for Privilege Escalation, and Exploitation for Client Execution.
The reason for this may be the fact that these are popular attack techniques and therefore were expressed by different researchers.
The table also presents the number of adversary groups mapped to each Technique, i.e., the number of groups that used these Techniques (and their Sub-techniques), and the number of software tools (used to conduct attacks) mapped to each Technique.
The mapping is based on ATT\&CK Enterprise v9.0.

As can be seen in the table, the average number of adversary groups using each Enterprise Technique is 13, and the average number of software tools using each Enterprise Technique is 31.
We can see that there are some expressed Techniques where the number of mapped groups and software tools is much higher than the average, e.g., Command and Scripting Interpreter, File and Directory Discovery, Process Injection and Phishing.
The number of groups and software tools using each Technique can be used to prioritize the Techniques to express.
The number of Group-Technique mappings (i.e., the number of adversary groups using each Technique, including Sub-techniques) for all Enterprise Techniques is 2,390; for expressed Techniques it is 811, which represents 34\% of Group-Technique mappings.
This percentage is much higher than the percentage of the expressed Techniques, which is 20\%.
This indicates that expressed Techniques are the more popular techniques used by adversaries.
The table also presents the papers with SIRs expressing each Technique, the number of times these papers have been cited, and the average number of citations per paper.

Figure~\ref{fig:CoveredTechniquesMatrix} (in the appendix) presents the Enterprise Techniques expressed by MulVAL rules as a matrix.

\begin{table*}[htbp]
    \centering
    \caption{Popularity analysis of expressed Techniques }
    \scriptsize
    \begin{tabular}{| m{17em} | c | c | c | c | c | c | c |} 
    \hline
    \textbf{Techniques} & \rotatebox{90}{\textbf{No. of Expressing SIRs}} & \rotatebox{90}{\textbf{No. of Using Groups}} & \rotatebox{90}{\textbf{No. of Using SW Tools}}  & \rotatebox{90}{\textbf{No. of Papers}} & \rotatebox{90}{\textbf{Papers with Expressing SIRs}} & \rotatebox{90}{\textbf{No. of Citations}} & \rotatebox{90}{\textbf{Average No. of Citations}}\\
    \hline
    External Remote Services & 1 & 17 & 4 & 1 & \cite{ou2011quantitative} & 50 & 50\\
    \hline
    Phishing & 1 & 98 & 31 & 1 & \cite{ou2005logic} & 57 & 57\\
    \hline
    Command and Scripting Interpreter & 1 & 185 & 351 & 1 & \cite{almohri2015security} & 47 & 47\\
    \hline
    Shared Modules & 1 & 0 & 11 & 1 & \cite{stan2020extending} & 6 & 6\\
    \hline
    System Services & 1 & 11 & 35 & 1 & \cite{inokuchi2019design} & 3 & 3\\
    \hline
    Abuse Elevation Control Mechanism & 1 & 9 & 30 & 1 & \cite{ou2005logic} & 57 & 57\\
    \hline
    File and Directory Discovery & 1 & 34 & 170 & 1 & \cite{dong2016right} & 0 & 0\\
    \hline
    Exploitation of Remote Services & 1 & 5 & 9 & 1 & \cite{sun2014inferring} & 9 & 9\\
    \hline
    Remote Service Session Hijacking & 1 & 0 & 1 & 1 & \cite{inokuchi2019design} & 3 & 3\\
    \hline
    Exfiltration Over Physical Medium & 1 & 2 & 5 & 1 & \cite{dong2016right} & 0 & 0\\
    \hline
    Disk Wipe & 1 & 5 & 6 & 1 & \cite{khakpour2019towards} & 4 & 4\\
    \hline
    Drive-by Compromise & 2 & 21 & 6 & 2 & \cite{ou2011quantitative} \cite{mccormack2020security} & 51 & 26\\ 
    \hline
    Process Injection & 2 & 24 & 112  & 2 & \cite{inokuchi2019design} \cite{sun2014inferring} & 12 & 6\\
    \hline
    Credentials from Password Stores & 2 & 31 & 73 & 1 & \cite{mccormack2020security} & 1 & 1\\
    \hline
    Exploitation for Credential Access & 2 & 0 & 0 & 1 & \cite{ou2005logic} & 57 & 57\\
    \hline
    Steal Application Access Token & 2 & 1 & 0 & 1 & \cite{acosta2016augmenting} & 8 & 8\\
    \hline
    Exploit Public-Facing Application & 3 & 14 & 3 & 3 & \cite{ou2005logic} \cite{inokuchi2019design} \cite{mccormack2020security} & 61 & 20\\
    \hline
    Valid Accounts & 3 & 47 & 13  & 2 & \cite{ou2005logic} \cite{mccormack2020security} & 58 & 29\\
    \hline
    Remote Services & 3 & 52 & 42 & 3 & \cite{inokuchi2019design} \cite{acosta2016augmenting} \cite{sun2014inferring} & 20 & 7\\
    \hline
    Exfiltration Over Alternative Protocol & 3 & 9 & 17 & 1 & \cite{zhou2020security} & 0 & 0\\
    \hline
    Data Manipulation & 3 & 4 & 4 & 2 & \cite{inokuchi2019design} \cite{zhou2020security} & 3 & 2\\
    \hline
    Endpoint Denial-of-Service & 3 & 1 & 2 & 2 & \cite{stan2020extending} \cite{zhou2020security} & 6 & 3\\
    \hline
    User Execution & 4 & 86 & 49 & 3 & \cite{ou2005logic} \cite{stan2020extending} \cite{mccormack2020security} & 64 & 21\\
    \hline
    OS Credential Dumping & 5 & 69 & 58 & 3 & \cite{ou2005logic} \cite{zhou2020security} \cite{mccormack2020security} & 58 & 19\\
    \hline
    Password Policy Discovery & 5 & 3 & 5 & 1 & \cite{stan2020extending} & 6 & 6\\
    \hline
    Steal or Forge Kerberos Tickets & 6 & 4 & 7 & 4 & \cite{ou2005logic} \cite{acosta2016augmenting} \cite{zhou2020security} \cite{mccormack2020security} & 66 & 17\\
    \hline
    Exfiltration over Other Network Medium & 6 & 0 & 1 & 1 & \cite{mccormack2020security} & 1 & 1\\
    \hline
    Data from Network Shared Drive & 7 & 6 & 4 & 3 & \cite{ou2005logic} \cite{stan2020extending} \cite{govindavajhala2006formal} & 65 & 22\\
    \hline
    Data from Local System & 9 & 28 & 60 & 4 & \cite{ou2005logic} \cite{inokuchi2019design} \cite{mccormack2020security} \cite{govindavajhala2006status} & 64 & 16\\
    \hline
    Network Denial of Service & 9 & 1 & 1 & 3 & \cite{stan2020extending} \cite{zhou2020security} \cite{mccormack2020security} & 7 & 2\\
    \hline
    Exploitation for Client Execution & 11 & 28 & 11 & 5 & \cite{ou2005logic} \cite{stan2020extending} \cite{zhou2020security} \cite{mccormack2020security} \cite{govindavajhala2006windows} & 102 & 20\\
    \hline
    Network Sniffing & 11 & 6 & 9 & 3 & \cite{stan2020extending} \cite{acosta2016augmenting} \cite{mccormack2020security} & 15 & 5\\
    \hline
    Exploitation for Privilege Escalation & 13 & 11 & 10 & 8 & \cite{ou2005logic} \cite{stan2020extending} \cite{ou2011quantitative} \cite{almohri2015security} \cite{sun2014inferring} \cite{zhou2020security} \cite{govindavajhala2006formal} \cite{khakpour2019towards} & 175 & 22\\
    \hline
    Man-in-the-Middle & 17 & 3 & 5 & 3 & \cite{stan2020extending} \cite{acosta2016augmenting} \cite{mccormack2020security} & 15 & 5\\
    \hline
    \hline
    Total &  & 811 & 1,145 &  & &  & \\
    \hline
    \end{tabular}
    \label{table:ExpressedTechniquesAnalysis}
\end{table*}

As an example of an expressed Technique, the Endpoint Denial of Service (DoS) Technique expressed by SIRs is presented in Listing~\ref{lst:dosexpression}:
\begin{lstlisting}[basicstyle=\ttfamily\footnotesize,language=Python,label=lst:dosexpression, caption=Endpoint DoS Technique expressed with SIRs]
dos(Principal, Host) :-
    localAccess(Principal, Host, User),
    localService(Host, Prog, User),
    vulHost(Host, VulID, Prog, localExploit, dos),
    malicious(Principal).
dos(Principal, DstHost) :-
    malicious(Prin),
    l2Access(Prin, SrcHost, DstHost, Prot, BusID, bus).
systemDown(Host) :-
    execCode(Host, _Perm2),
    vulExists(Host, _, SW, localExploit, Overuse),
    misuseAction(Overuse).
\end{lstlisting}

\section{Related Work}
Several previous studies performed surveys of different attack generation tools.
Yi et al.~\cite{yi2013overview} surveyed and analyzed attack graph generation and visualization technology, and compared several academic and commercial attack graph generation tools.
Barik et al.~\cite{barik2016attack} presented a consolidated view of major attack graph generation and analysis techniques.
In an extensive survey of relevant papers, Haque et al.~\cite{haque2017evolutionary} summarized the different approaches to attack modeling, i.e., attack graphs and attack trees.
Hong el al.~\cite{hong2017survey} discussed the current state of graphical security models in terms of four phases: generation, representation, evaluation, and modification.
Garg et al.~\cite{garg2018systematic} conducted a literature review, focusing on the generation and analysis of attack graphs.
He et al.~\cite{he2019unknown} surveyed unknown vulnerability risk assessment based on directed graph models and classified their security metrics.
By analyzing more than 180 attack graphs and attack trees, Lallie et al.~\cite{lallie2020review} presented empirical research aimed at identifying how attack graphs and attack trees present cyber attacks in terms of their visual syntax.
None of the papers mentioned above surveyed MulVAL attack graph generation extensions.

Many studies mapped attack entities, such as malware,  CVE, and CTI, to MITRE ATT\&CK, as the defacto standard for cyber threat modeling.
Oosthoek et al.~\cite{oosthoek2019sok} mapped Windows malware families to the ATT\&ACK framework.
Legoy~\cite{legoy2019retrieving} evaluated different multi-label text classification models to retrieve TTPs from textual sources, based on the ATT\&CK framework, and developed a tool for extracting ATT\&CK Tactics and Techniques from cyber threat reports to a structured format.
Aghaei et al.~\cite{aghaei2019threatzoom} suggested using machine learning, deep learning, and natural language processing to map CVE to CAPEC and ATT\&CK automatically, and found the appropriate mitigation for each CVE.
By mapping the MITRE ATT\&CK Matrix to the NIST cyber security framework, Kwon et al.~\cite{kwon2020cyber} offered approaches and practical solutions to cyber threats.
Purba et al.~\cite{purba2020word} defined a cyber-phrase embedding model to map CTI texts to the ATT\&CK ontology.
They created an ontology based on MITRE ATT\&CK, by integrating 2,236 attack patterns associated with ATT\&CK Tactics and Techniques.
Lee et al.~\cite{lee2020fileless} analyzed 10 selected cyber attacks in which fileless techniques were utilized and mapped the attacks to ATT\&CK Techniques.
However, none of these works mapped MulVAL interaction rules to MITRE ATT\&CK Techniques.

To the best of our knowledge, we are the first to survey all MulVAL extensions, and map all of the MulVAL interaction rules to MITRE ATT\&CK Techniques.

\section{Summary}
AGs in general and MulVAL in particular are important tools for network risk assessment and cybersecurity improvement.
For providing a comprehensive risk assessment of an organization’s network, attack graphs should be able to present as many attack scenarios as possible.
The main security goal of this paper is assessing the coverage of attack scenarios supported by a popular logical AG generation framework. 

\textbf{Insights.}
Our main insights are: 
1) MulVAL is the most commonly used attack graph generation framework in academic research. 
2) MulVAL interaction rules can be mapped to ATT\&CK Tactics and Techniques. 
3) Today MulVAL rules cover less than a quarter of the ATT\&CK Techniques; therefore, it cannot be considered a reliable risk assessment tool yet. 
4) There is a need for AGs with a complete and up-to-date coverage of attack scenarios. 

\textbf{Main contributions.}
Since MulVAL was introduced in 2005, interaction rules have been added to represent additional attack scenarios.
In this paper, we surveyed the 938 academic publications mentioning MulVAL and identified 38 papers extending MulVAL.
To improve the usefulness of MulVAL, we provide the list of all MulVAL interaction rules, that can work together to enable broader risk assessment.
To evaluate the extent to which MulVAL rules are able to represent different attacks, we mapped all of the MulVAL rules to MITRE ATT\&CK Techniques and summarized the attack coverage capabilities provided by the MulVAL rules.

Mapping between the most commonly used attack graph generation tools, such as MulVAL, and the MITRE ATT\&CK threat model will enable security administrators to handle more realistic attack scenarios.
A clear understanding of an existing network's strength against different types of TTPs is critical, and the simulation of MITRE-based attack scenarios enables such understanding.
For example, this can help security administrators decide which defensive measures to implement.

\textbf{Main challenges.} 
The main challenge we faced while conducting this survey was the lack of a standard terminology across the published MulVAL extensions. 
For example, the meaning of \textbf{User} differs among the published papers -- it may relate to the configured host account or to the logical user principal.
We also found that since the MulVAL extensions were generated by different researchers, there are some duplicate rules.
In addition, the MulVAL related articles do not relate the proposed interaction rules to MITRE TTPs. 
Mapping the rules to the most appropriate ATT\&CK Techniques posed an additional challenge.

\textbf{Future work.} In future work we intend to normalize the MulVAL rules, removing interaction rules that were defined more than once with different names, different parameter names, or a different order of parameters.
In addition, we plan to propose a methodology for expressing arbitrary ATT\&CK Techniques using MulVAL interaction rules.
A grand challenge would be modeling the entire known attack scenario, e.g., all the ATT\&CK Techniques, to interaction rules.
This will enhance MulVAL's ability to provide realistic network risk assessment.
The next milestones on the MulVAL development road-map may be MITRE ATT\&CK Mobile and ICS Techniques.
Finally, this MulVAL extensions development would highly benefit from automation in the interaction rule generation process.

\bibliographystyle{IEEEtran}
\bibliography{References}

%\newpage

\section{Biography Section}
\begin{IEEEbiography}[{\includegraphics[width=1in,height=1.25in,clip,keepaspectratio]{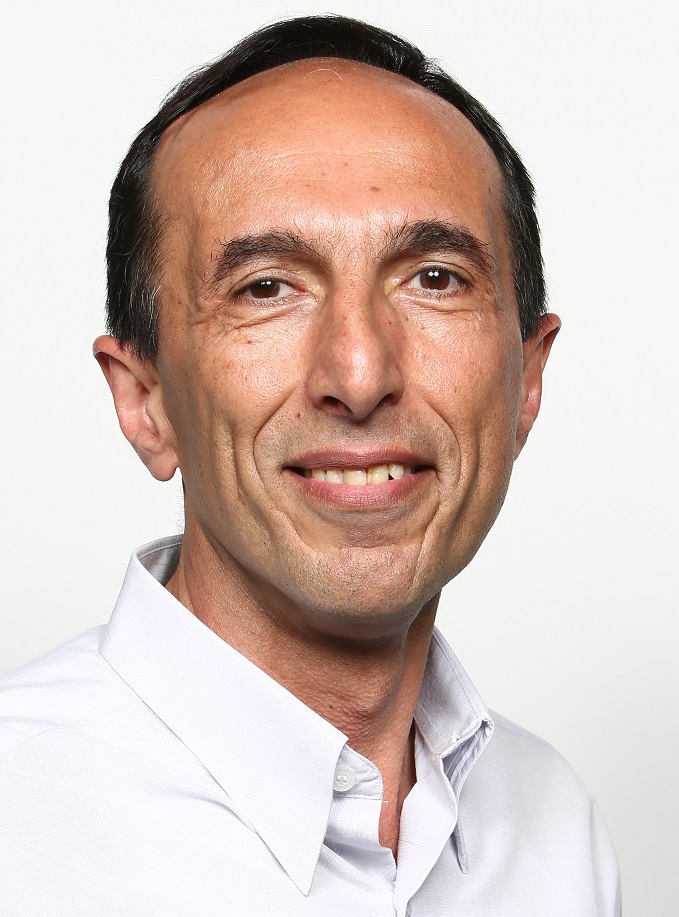}}]{David Tayouri} holds BSc and MSc degrees in computer science and is a doctoral student in the Department of Software and Information Systems Engineering at Ben-Gurion University of the Negev.
His main areas of interest are cyber security and attack graphs.
\end{IEEEbiography}
\begin{IEEEbiography}[{\includegraphics[width=1in,height=1.25in,clip,keepaspectratio]{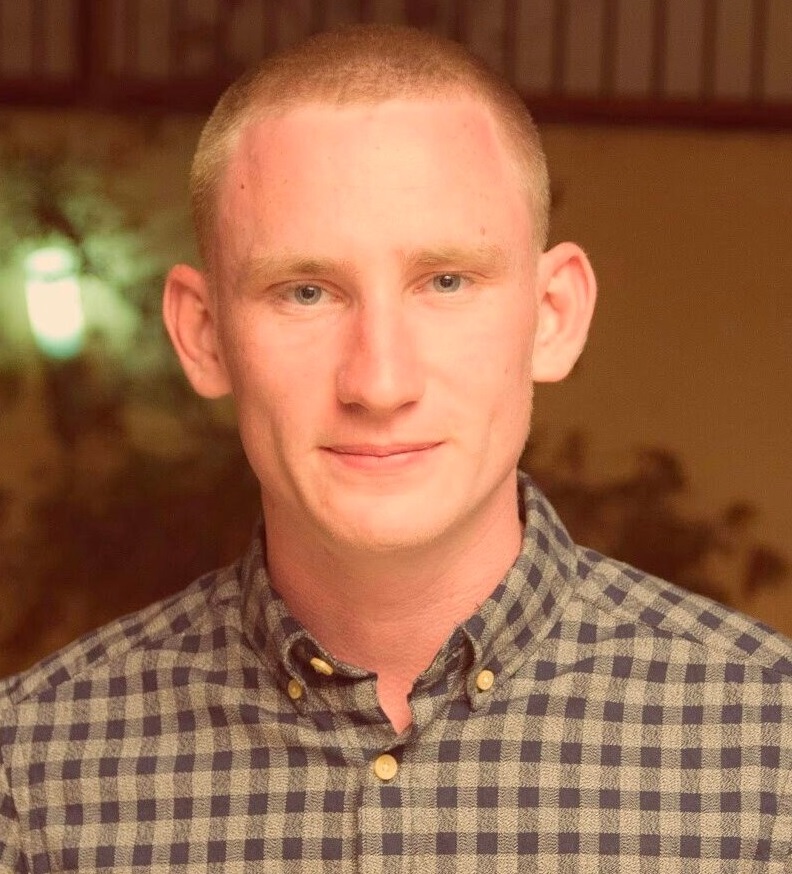}}]{Nick Baum} holds BSc degree in computer science and is a master student in the Department of Software and Information Systems Engineering at Ben-Gurion University of the Negev.
\end{IEEEbiography}
\begin{IEEEbiography}[{\includegraphics[width=1in,height=1.25in,clip,keepaspectratio]{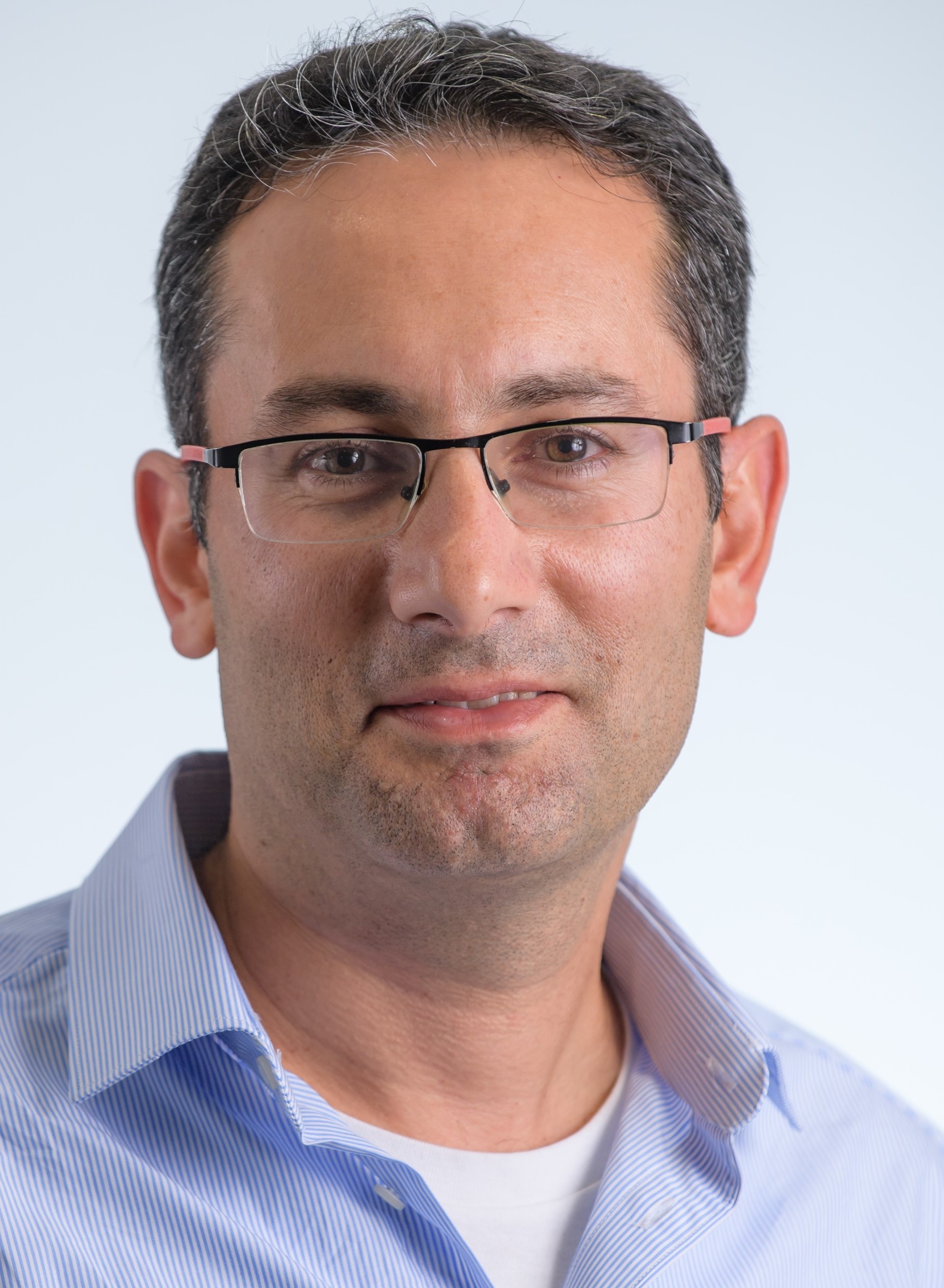}}]{Asaf Shabtai} is a professor in the Department of Software and Information Systems Engineering at Ben-Gurion University of the Negev.
His main areas of interest are computer and network security, machine learning, and security of the IoT, smart mobile devices, and operational technology (OT) systems.
\end{IEEEbiography}
\begin{IEEEbiography}[{\includegraphics[width=1in,height=1.25in,clip,keepaspectratio]{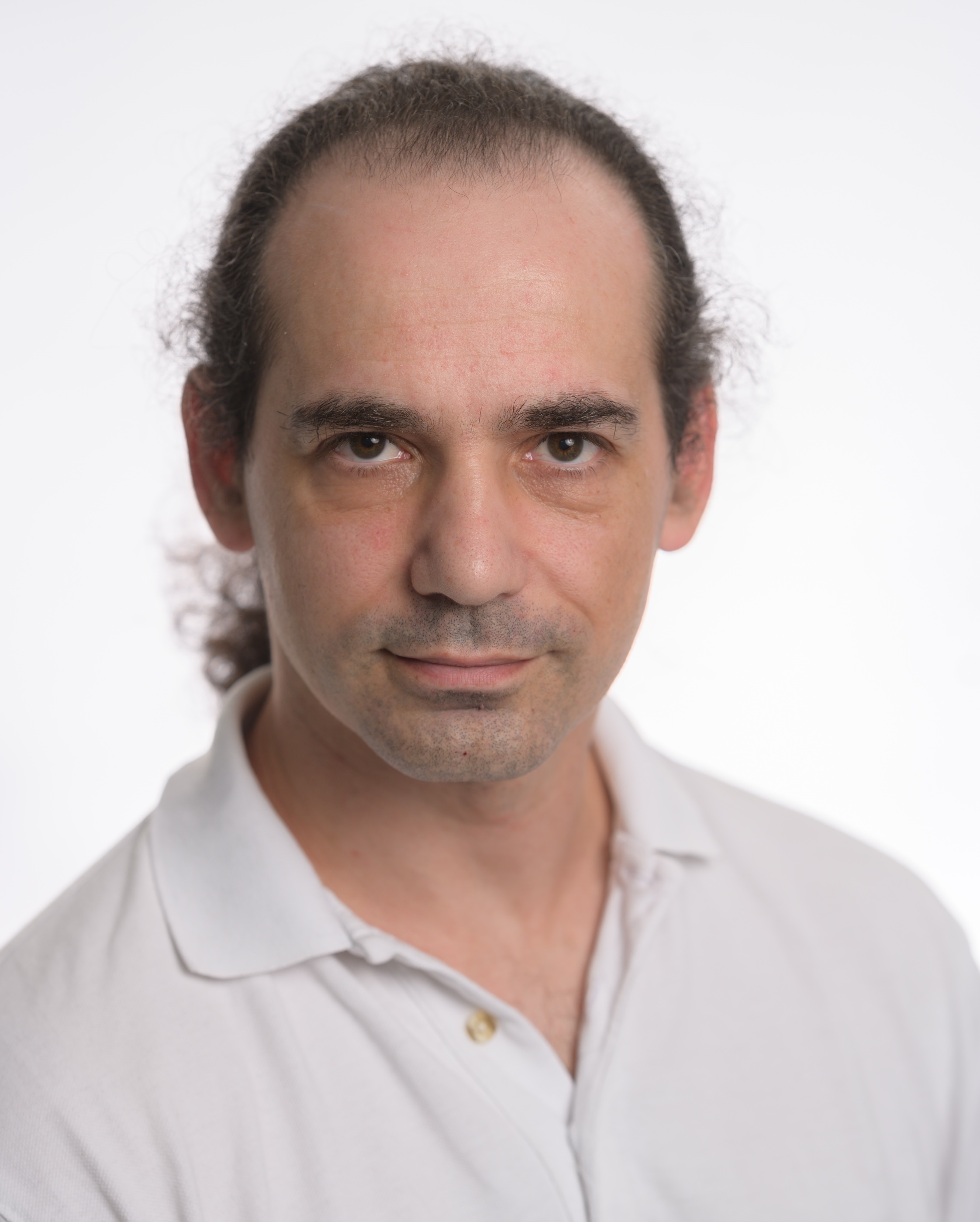}}]{Rami Puzis}
is a faculty member at the Department of Software and Information Systems Engineering at Ben-Gurion University.
Rami has graduated BSc in Software Engineering and MSc and PhD in Information Systems Engineering. All three degrees were received with honors from Ben-Gurion University.
He was a post-doctoral research associate in the Lab for Computational Cultural Dynamics, University of Maryland.
His main research interests include network analysis with applications to security, social networks, communication, and biology.
\end{IEEEbiography}

\vfill

\end{document}